%% file: main.tex
\documentclass{INTERSPEECH2023}

\makeatletter
\def\bstctlcite{\@ifnextchar[{\@bstctlcite}{\@bstctlcite[@auxout]}}
\def\@bstctlcite[#1]#2{\@bsphack
  \@for\@citeb:=#2\do{%
    \edef\@citeb{\expandafter\@firstofone\@citeb}%
    \if@filesw\immediate\write\csname #1\endcsname{\string\citation{\@citeb}}\fi}%
  \@esphack}
\makeatother

\usepackage[table,xcdraw]{xcolor}
\usepackage{multirow}

\interspeechcameraready

\title{Some voices are too common: Building fair speech recognition systems using the Common Voice dataset}
\name{Lucas Maison$^{1, 2}$, Yannick Estève$^1$}
\address{
  $^1$Laboratoire Informatique d'Avignon, Avignon, France\\
  $^2$ Thales SIX, Multimedia Lab, Gennevilliers, France}
\email{lucas.maison@alumni.univ-avignon.fr}

\begin{document}
\bstctlcite{IEEEexample:BSTcontrol}

\maketitle
 
\begin{abstract}
Automatic speech recognition (ASR) systems become increasingly efficient thanks to new advances in neural network training like self-supervised learning. However, they are known to be unfair toward certain groups, for instance, people speaking with an accent. In this work, we use the French Common Voice dataset to quantify the biases of a pre-trained wav2vec~2.0 model toward several demographic groups. By fine-tuning the pre-trained model on a variety of fixed-size, carefully crafted training sets, we demonstrate the importance of speaker diversity. We also run an in-depth analysis of the Common Voice corpus and identify important shortcomings that should be taken into account by users of this dataset.
\end{abstract}

\input{sections/1_intro}
\input{sections/2.dataset}
\input{sections/3.experiments}
\input{sections/4.results}

\input{sections/5.conclusion}

%\section{Acknowledgements}
%The authors would like to thank ISCA and the organising committees of past INTERSPEECH conferences for their help and for kindly providing the previous version of this template.

\newpage

\bibliographystyle{IEEEtran}
\bibliography{mybib}

\end{document}

%% file: sections/1_intro.tex
\section{Introduction}
\label{sec:intro}

% intro and related work

Recent years have seen great advances in automatic speech processing, thanks to the generalization of new techniques for building speech representations using neural networks, including self-supervised learning~(SSL)~\cite{ssl_review, w2v2, hubert, Chen2021WavLM}, self-training~(PT)~\cite{Park2020NST}, or a combination of both~\cite{Xu2021STandPT, Zhang2020w2vconformer, chung2021w2vBERT, Zhang2021BigSSL}. SSL and PT are used to leverage huge amounts of unannotated data to build powerful speech representations that can be used to tackle various downstream tasks like automatic speech recognition~(ASR) or speaker identification. To this end, models pre-trained with a SSL approach are then fine-tuned on a target domain. Although it has been shown that speech recognition models can be trained effectively in a fully-supervised manner~\cite{gulati20conformer, whisper}, to this day self-supervision remains popular for building powerful ASR systems~\cite{USM}.

There is a long history of bias in speech recognition~\cite{Tatman2017, Koenecke2020}. Despite their state-of-the-art performance, there is growing evidence that even the most recent models are not robust to domain shifts~\cite{Hsu2021, gomez2023ATC} or data bias~\cite{meng2022dont, asr4real}. Examples of domain shifts include acoustic conditions~\cite{Zhu2022noiserobust}, vocabulary and grammar~\cite{gomez2023ATC}, and speech style~(read speech, spontaneous speech)~\cite{Hsu2021, Likhomanenko2021}, whereas models can be biased towards specific demographic attributes, like gender~\cite{asr4real, Boito2022}, age~\cite{draft}, accent~\cite{asr4real, Koenecke2020, accent_survey}, or prosody~(speech rate)~\cite{meng2022dont}.

Various methods for dealing with domain shifts have been referenced in the literature, like using data augmentation~\cite{Fukuda2018}, adding adapters to the backbone model~\cite{draft}, exploiting both spectral and SSL features~\cite{berrebbi22_interspeech}, or increasing the amount of pre-training data~\cite{Hsu2021, Likhomanenko2021}.

In this work, we show how pre-trained models' performance is biased toward certain demographics, and investigate whether fine-tuning on carefully designed training sets can counterbalance these biases. Our work is closely related to~\cite{Boito2022}, where they pre-train gender-specific models to investigate gender bias. In~\cite{Hsu2021, Likhomanenko2021}, authors pre-train models using various datasets and data sizes but do not investigate the role of fine-tuning, nor do they break down results by demographics. Fairness in ASR has been investigated~\cite{casual_conversations, trinh22_interspeech}, however, these works do not analyze the impact of biased fine-tuning sets. In~\cite{asr4real}, authors fine-tune SSL models on different domains and evaluate gender and accent biases, but they do not control for the size of the datasets. In this work, we investigate the impact of fine-tuning a SSL model on a single domain, controlling for data size and demographic biases. We also provide a critical analysis of the French Common Voice dataset. To the best of our knowledge, this is the first work investigating in-depth the impact of the fine-tuning step on the demographic biases of a wav2vec~2.0 model.

%% file: sections/2.dataset.tex
\section{Common Voice dataset}
\label{sec:cv_dataset}

Common Voice~\cite{commonvoice:2020} is a massive crowd-sourced multilingual corpus of read speech. It is maintained by Mozilla and is freely available online\footnote{\url{commonvoice.mozilla.org/datasets}}. We use the French subset of the Common Voice~12.0 corpus, which is the latest available version at the time of writing. It consists of more than 1,000 hours of speech~($\approx$ 760,000 utterances) spoken by more than 17,000 different speakers. Anyone can contribute to expanding the dataset by recording its voice or validating other users' recordings. Interested readers can refer to the article~\cite{commonvoice:2020} for more details on the dataset creation process.

\noindent Utterances are partitioned into three splits (validated, invalidated, and other) depending on their validation status. The validated split itself contains three disjoint subsets (train, dev, test), each with its unique speakers and sentences. Statistics for each split are reported in Table~\ref{tab:dataset}.

\input{tables/dataset}

We can see that more than 92,000 utterances are yet to be validated or have been deemed \textit{invalid} by the community. While a lot of these rejects are justified~(because of inaudible audio, wrong sentence, multiple voices), manual inspections of the data show that some samples were wrongfully rejected. Conversely, some samples are deemed \textit{valid} when they are not. We explore the influence of samples' quality on ASR performance in the following sections.

\subsection{Demographic bias}

\begin{figure}[h]
    \begin{center}
    \includegraphics[width=\columnwidth]{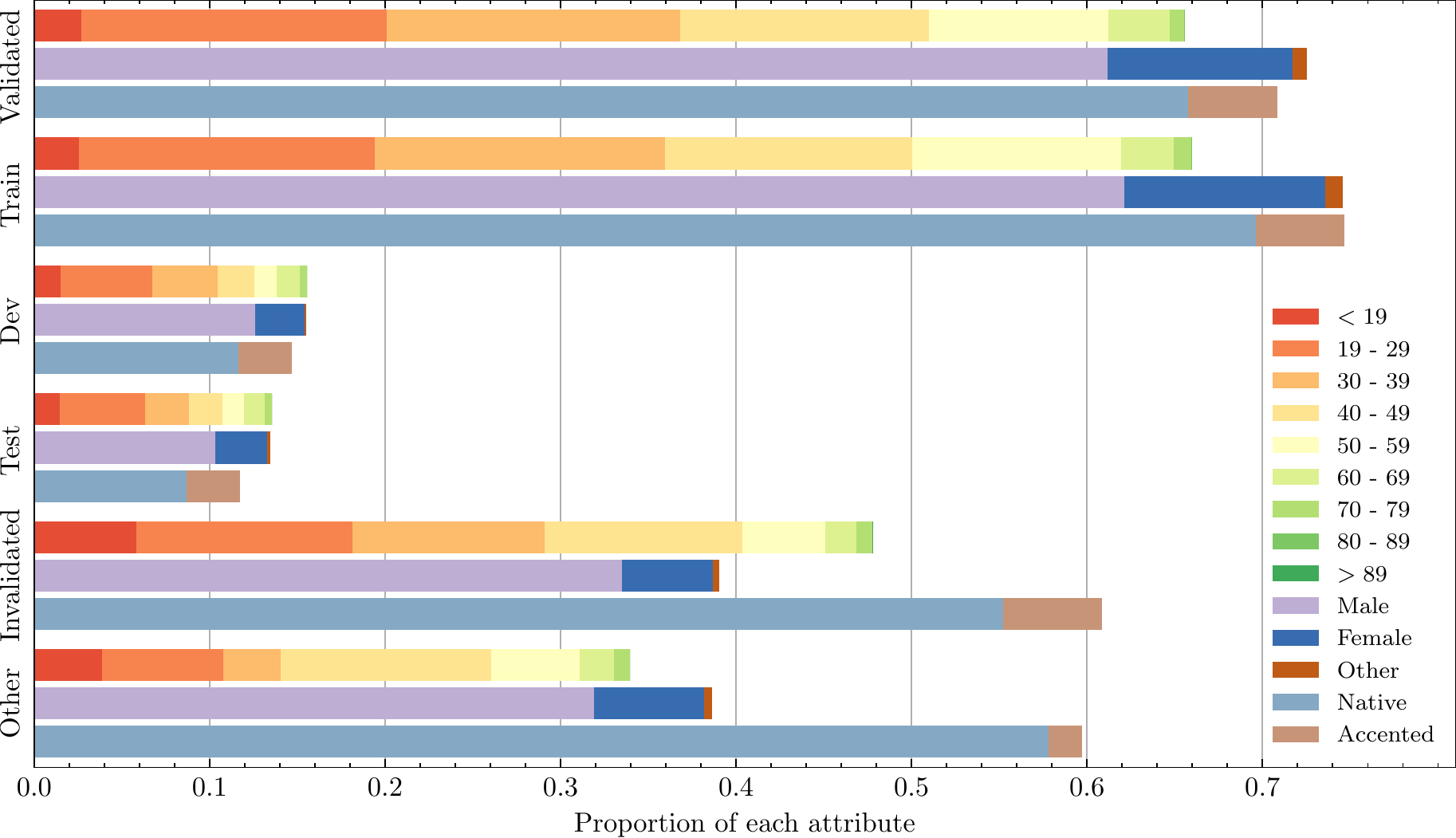}
    \caption{Distribution of each demographic attribute for the different data splits. Contributors may opt out of sharing demographic data; hence the proportions not summing to $1$.}
    \label{fig:demographics}    
    \end{center}
\end{figure}

One important feature of Common Voice is that users can self-report their demographic information on their profile; this information is then shipped alongside the dataset. Demographic data comprises age category, gender, and accent. We report the distributions of each of these attributes in Figure~\ref{fig:demographics}.

\noindent We can clearly see the data imbalances for each attribute. Taking the train set as an example, we observe that 83\% of all labeled utterances (62\% of the total) come from male speakers, whereas only 15\% (11\% of the total) come from female speakers. The discrepancy is even higher for the accent attribute, and we observe strong differences between age classes. Although these can be partially explained by the French population pyramid, the gender gap is more difficult to justify and could motivate further investigations.

We expect that these data biases will have repercussions on models fine-tuned on this trainset (or on random subsets of it). We test this hypothesis in the following sections.

\subsection{Speaker bias}

\begin{figure}[h]
    \begin{center}
    \includegraphics[width=\columnwidth]{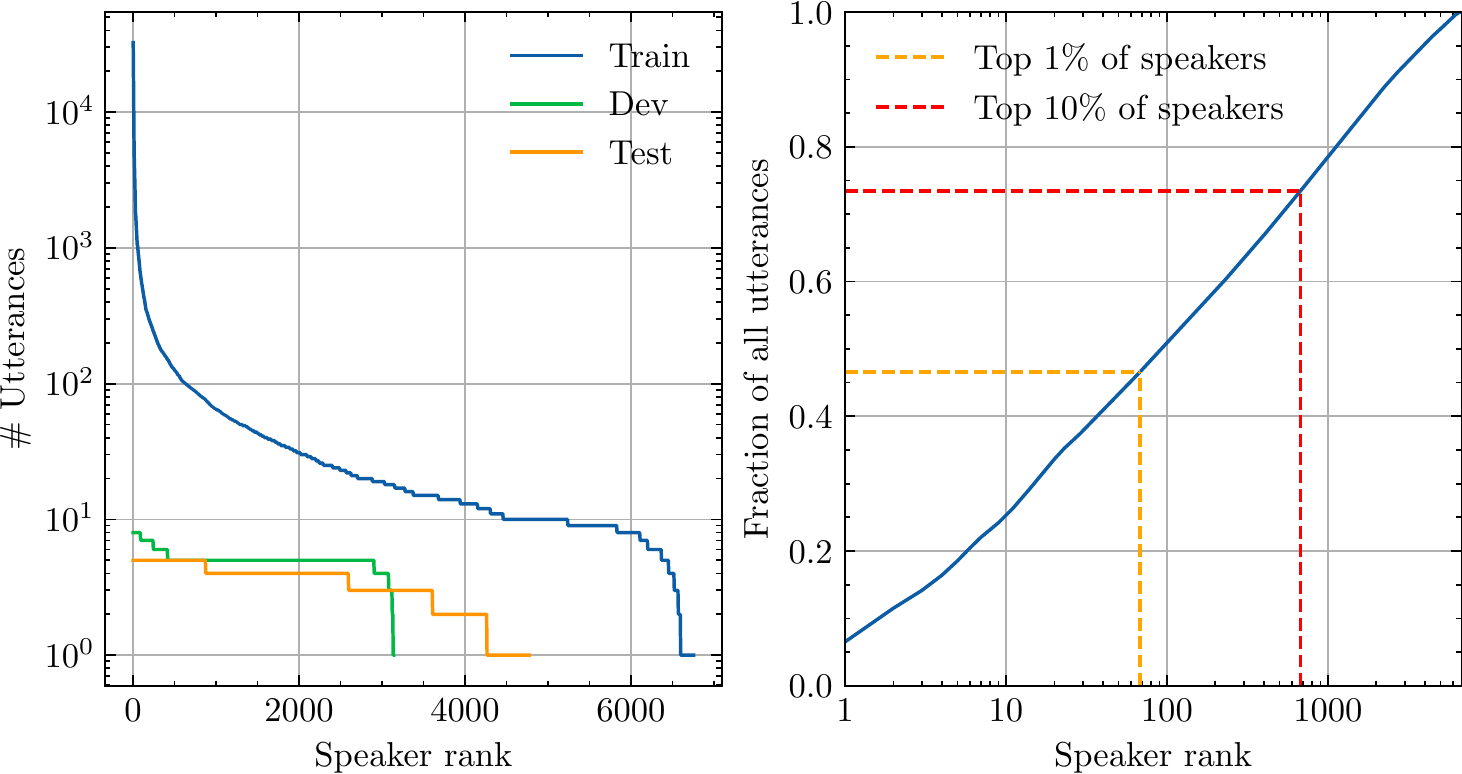}
    \caption{Left: Distribution of utterances between speakers for train/dev/test splits. Right: Cumulative distribution of utterances for the train split. Speaker rank is the rank in the list of speakers sorted by decreasing number of utterances.}
    \label{fig:common_voice}    
    \end{center}
\end{figure}

There are no limits on a user's contributions to the dataset ; hence, it is expected that some users will record more sentences than others. To get a better sense of whether this is the case, we plot the number of utterances by speaker~(Figure~\ref{fig:common_voice}, left). As we can see, the dev and test sets are balanced, with no more than 8 utterances per speaker. On the other hand, the train set is heavily unbalanced, with some speakers having several orders of magnitude more utterances than others. To better visualize this distribution, we plot the cumulative distribution of utterances on the right panel of Figure~\ref{fig:common_voice} (note the logarithmic axis). We observe that 7\% of all utterances were spoken by a single user, while the top ten speakers represent 25\% of all utterances. One-half and three-quarters of all utterances are spoken by only 1\% and 10\% of all the speakers, respectively. The training set is therefore clearly biased toward certain speakers. This could be a concern when taking sub-samples of this set, and indeed we find that random samples of size 50,000 contain 5,275 speakers on average, a 22\% decrease. We explore this problem further in the following sections.

\section{Training sets}
\label{sec:train_sets}

In order to answer the questions raised in section~\ref{sec:cv_dataset}, we build various training sets using different strategies for sub-sampling the original train set. To ensure our comparisons are fair, each training set contains exactly 50,000 utterances~($\approx 71 \pm 3$ hours of audio data). Choosing such a reduced data size has several benefits. It makes the training faster and consumes less energy. It also allows us to create fully biased train sets~(for instance, 100\% female speech, see section~\ref{subsec:demographic_train_sets}) even for low-resource demographic subgroups.

\noindent We begin by creating \emph{reference} train sets, which are simply random sub-samples of the train split. These sets will provide us with baseline performance. Next, we create a \emph{"bad quality"} set using invalidated samples, and two \emph{"high quality"} sets containing only samples with no downvotes or samples with the most upvotes, respectively. These sets will be useful to determine the relevance of the samples' quality.

\subsection{Demographic train sets}
\label{subsec:demographic_train_sets}
To measure the impact of demographic biases, we build a series of train sets whose characteristics are summarized in Table~\ref{tab:training_sets}. For each of the three demographic attributes, we create sets voluntarily biased towards certain values of that attribute. Whenever possible, we try to affect the same number of speakers to the different train sets related to the same attribute (see for instance how "male only", "female only" and "mixed genders" all have 598 speakers). We do so in order to eliminate a possible confusion factor. Finally, we also create a \emph{"high diversity"} set which is optimal in terms of demographic distribution: each attribute has its values equally represented in the train set.

\input{tables/training_sets}

\subsection{Varying the number of speakers}
\label{subsec:speakers}

We suspect that the number of speakers in a training set has a strong influence on the final recognition performance. To test this hypothesis independently of the data size, we create several sets of size 50,000 with 2\footnote{We use all the data from the top speaker (32,504) and complete with the second top speaker.}, 10, 100, 1,000, and 6,756\footnote{The train split contains 6,756 unique speakers} speakers.

\noindent We also want to test whether it is more important to include more samples or to diversify the speakers.  To do so, we create two additional sets, \emph{Small} and \emph{Medium}, with 6,756 speakers each, and respectively one and three utterances per speaker.

%% file: tables/dataset.tex
\begin{table}[th]
  \caption{Statistics for the different splits of the dataset. Duration is hours:minutes.}
  \label{tab:dataset}
  \centering
\begin{tabular}{l|ccc}
\textbf{Split}      & \textbf{\# Speakers} & \textbf{\# Utterances} & \textbf{Duration} \\ \hline
validated           & 16140                & 666754                 & 915:42                    \\
$\rightarrow$ train & 6756                 & 499535                 & 718:38                    \\
$\rightarrow$ dev   & 3140                 & 16104                  & 25:46                     \\
$\rightarrow$ test  & 4774                 & 16104                  & 26:14                     \\ \hline
invalidated         & 9365                 & 58418                  & 91:56                     \\
other               & 845                  & 34255                  & 47:25                    
\end{tabular}
  
\end{table}

%% file: tables/training_sets.tex
\begin{table}[th]
  \caption{Description of the content of the demographic train sets, along with the number of speakers. $\dagger$This train set contains repeated utterances to compensate for a lack of data.}
  \label{tab:training_sets}
  \centering
\begin{tabular}{l|c|c}
\textbf{Trainset} & \textbf{Description}                                                                  & \textbf{Speakers} \\ \hline
Reference         & \begin{tabular}[c]{@{}c@{}}A random subsample\\ of the train split\end{tabular}       & $\approx$ 5,275   \\ \hline
Female only       & 100\% female                                                                          & 598               \\
Male only         & 100\% male                                                                            & 598               \\
Mixed genders     & 50\% female, 50\% male                                                                & 598               \\ \hline
Youngs only       & 25\% teens, 75\% twenties                                                             & 1,303             \\
Middle-age only   & 50\% thirties, 50\% fourties                                                          & 1,280             \\
Seniors only      & \begin{tabular}[c]{@{}c@{}}60\% fifties, 30\% sixties,\\ 10\% seventies+\end{tabular} & 510               \\
Mixed ages        & \begin{tabular}[c]{@{}c@{}}10\% seventies+, 15\% for\\ each other age\end{tabular}    & 1,285             \\ \hline
Native only       & 100\% "French of France"                                                              & 432               \\
Accented only     & 100\% other accents$\dagger$                                                          & 432               \\
Mixed accents     & \begin{tabular}[c]{@{}c@{}}50\% native,\\ 50\% accented speech\end{tabular}           & 432               \\ \hline
High diversity    & \begin{tabular}[c]{@{}c@{}}Equal distribution among\\ attributes\end{tabular}         & 1,079            
\end{tabular}
\end{table}

%% file: sections/3.experiments.tex
\section{Experiments}

\subsection{Data preparation}
\label{subsec:data_preproc}

We keep the audio files as they are, simply converting them from \SI{48}{\kilo\hertz} MP3 to \SI{16}{\kilo\hertz} WAV format. We also filter out the few audio files longer than \SI{12}{\second} for efficiency purposes.

\noindent Regarding the text labels, we do extensive cleaning in order to obtain the most faithful transcriptions possible. First, we convert symbols like $\alpha$ and \$ to their written forms. Then we normalize non-Latin Unicode characters into their ASCII form while taking special care of keeping French-accented characters. Finally, we filter out all the remaining special characters and punctuation. While not perfect, we believe this procedure is a good heuristic for cleaning the transcriptions and validated it empirically by listening to corner cases.

\subsection{Model}
\label{subsec:model}

We use the Wav2Vec~2.0 model~\cite{w2v2} for all our experiments ; specifically, we use the LB-7K-large model from the \textit{LeBenchmark}~\cite{lebenchmark} initiative, which was pre-trained on 7,739 hours of French audio. Hyperparameters and architecture are identical to the ones first introduced in~\cite{w2v2}. Note that we use the \textit{large} architecture variant, which presents greater capacity~(317 million parameters). We use the \textit{LB-7K} variant of the models since previous work~\cite{maison2022promises, lebenchmark} has shown that for this task, models pre-trained using the greater quantity of audio performed best.

The pre-trained Wav2Vec~2.0 model acts as a speech encoder, which is fine-tuned for the ASR task together with an additional feed-forward network. This head network consists of three linear layers with 1,024 neurons. Each linear layer is followed by batch normalization and a leaky ReLU~\cite{Maas2013RectifierNI} activation function. We use dropout with $p=0.15$ between each linear layer.
\noindent At last, a final linear layer projects the output into token space, and log-softmax is applied to obtain probabilities of each token. We use 42 tokens, which represent individual characters.

\subsection{Training}
\label{subsec:training}

We use the \texttt{SpeechBrain}~\cite{speechbrain} toolkit for all our experiments. All our models are fine-tuned during 50 epochs using the CTC loss. We decided to freeze the convolutional layers of the transformer during fine-tuning since preliminary experiments taught us that this speeds up training by 15\% with little to no impact on performances. Adam~\cite{adam} and Adadelta~\cite{Zeiler2012ADADELTAAA} optimizers with learning rates $10^{-4}$ and $1.0$ are used to update the weights of the Wav2Vec~2.0 model and the additional top layers respectively. Learning rates are reduced at each epoch in which the validation loss does not improve.

\noindent During training, we apply on-the-fly data augmentation using the \texttt{Speechbrain} time-domain approximation of the SpecAugment~\cite{Park2019specaugment} algorithm: it disrupts audio speed, and randomly drops chunks of audio and frequency bands. We disable audio speed modification since we find it to destabilize training while providing little or no performance improvement.

For fine-tuning we use several different training sets, which were formed by sampling audio data following various protocols~(varying speaker diversity, demographics, speech quality, etc). We detail the formation of these training sets in section~\ref{sec:train_sets}. We use the official dev set~(validation set) for early stopping; this set is composed of 26 hours of audio, see Table~\ref{tab:dataset}.

Each model is trained on a single V100 GPU. Fine-tuning a model for 50 epochs on \SI{70}{\hour} of audio takes $\approx$ \SI{44}{\hour} of compute.

\subsection{Evaluation}
\label{subsec:evaluation}

We evaluate our trained models on the official test set, which stays identical for all the experiments;  this set is composed of 26 hours of audio, see Table~\ref{tab:dataset}. We use the Word Error Rate~(WER) as our test metric; lower is better. Note that we do not use any language model besides our end-to-end model.

%% file: sections/4.results.tex
\section{Results}
\label{sec:results}

\subsection{Does the quality of the samples matter?}

We evaluate our reference models on the test set and obtain a WER of $13.7 \pm 0.1$. We compare this result with our two \emph{"high quality"} models, which score 13.7 and 13.9 respectively. We deduce that the number of upvotes or downvotes of validated samples barely matters. On the other hand, our \emph{"bad quality"} model trained using invalidated samples scores 16.7, a significant drop in performance. However, it is quite noteworthy that a model fine-tuned on samples that were deemed unusable is still able to attain a reasonable ASR performance. This suggests that the invalidated split still contains a lot of usable audio, therefore the rejection process of Common Voice could be made less strict, which would contribute to increasing the dataset size.

\subsection{Can we reduce bias toward certain demographic groups?}

\input{tables/demographic_results}

For each model trained on the different train sets described in Table~\ref{tab:training_sets}, we evaluate its performance on the test set and break down results by demographic attributes (see Table~\ref{tab:demographic_results}). Surprisingly, we observe that models trained on biased train sets perform no better than the reference on the subgroup they are supposed to target. For instance, the model trained on the \emph{"female only"} set scores 16.9 on the female test subset, whereas the reference model scores 16.3. This trend holds true for all the subgroups for our analysis.

Overall, there are persistent biases toward some demographic subgroups: male speech is better recognized than female speech, young people's speech is badly recognized compared to the other age categories, and native speech is far better recognized than accented speech. We believe that these biases are rooted in the underlying pre-trained model; indeed, LB-7K-large~\cite{lebenchmark} has been pre-trained on a dataset that contains almost no accented data and is heavily biased toward male speech\footnote{See~\cite{lebenchmark}, Table~1. Note that although data from VoxPopuli~\cite{wang-etal-2021-voxpopuli} is not gender-annotated, it comes from the European Parliament which used to consist of a majority of men.}.

Finally, we observe that even our \emph{"high diversity"} model does not help to decrease subgroups biases. Although it seems to perform better than the reference on some subgroups, we find that these improvements are not statistically significant.

\subsection{Influence of the number of speakers}

\begin{figure}[h]
    \begin{center}
    \includegraphics[scale=0.8]{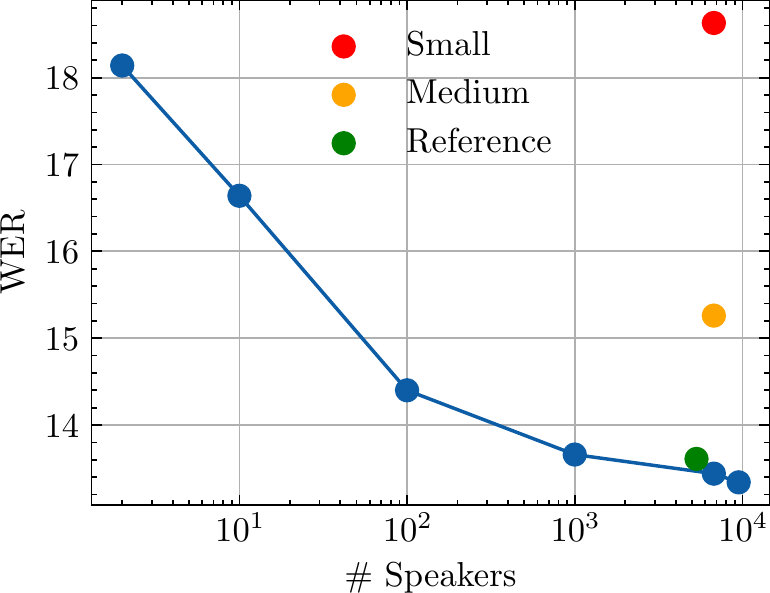}
    \caption{WER as a function of the number of speakers. Each point represents a different train set. Note that all train sets are of the same size (50,000 utterances) except \emph{Small} (6,756) and \emph{Medium} (19,918).}
    \label{fig:wer_spk}    
    \end{center}
\end{figure}

We evaluate the models fine-tuned on the train sets described in section~\ref{subsec:speakers} which have varying numbers of speakers. Results are shown in Figure~\ref{fig:wer_spk}, blue curve. We can see that the number of speakers (hence, the diversification of timbres, recording conditions, etc) in the training set has a clear impact on the WER. The higher the number of speakers, the lower the WER. When fine-tuning the model on a dataset containing more speakers than the reference, it obtains a test WER of 13.4, beating the score of the reference.

In an effort to gather even more speakers, we split the dev set (90-10 split) and use the largest part to augment our training set, increasing the number of different speakers to 9,500. We obtain a WER of 13.3 with a model trained on this set, which is consistent with our previous findings. Moreover, models trained on these sets obtain lower WER scores than the reference for each demographic subgroup, see Table~\ref{tab:best_results}. We conclude that researchers looking for the fairest fine-tuning dataset should focus on maximizing the number of speakers rather than maximizing demographic diversity.

Finally, models trained on a large number of speakers (6,756) but on smaller train sets have performance on par with models trained on more data but with a small number of speakers (see Figure~\ref{fig:wer_spk}). This highlights the importance of prioritizing speaker diversity over dataset size when collecting audio data.

\input{tables/best_results}

%\subsection{What if we fine-tune on more data?}

%% file: tables/demographic_results.tex
\begin{table*}[h]
\caption{Results of models trained on demographic-biased datasets. WER on the full test set is reported in the column \emph{test}. We also report the mean WER across demographic attributes. $\dagger$In this row we report the mean WER across our reference models. The best results in each column are shown in bold.}
\label{tab:demographic_results}
\centering
\begin{tabular}{l|c|cc|ccc|cc|c}
\textbf{Train set}  & \textbf{test} & \textbf{female}              & \textbf{male}                & \textbf{youngs}                       & \textbf{middle-aged}         & \textbf{seniors}             & \textbf{native}              & \textbf{accented}            & \textbf{mean} \\ \hline
reference $\dagger$ & \textbf{13.7} & 16.3                         & \textbf{14.3}                & 16.1                                  & 13.7                         & \textbf{13.4}                & \textbf{12.6}                & 17.4                         & \textbf{14.8} \\ \hline
female only         & 14.4          & \cellcolor[HTML]{EEEEEE}16.9 & \cellcolor[HTML]{EEEEEE}15.3 & 16.7                                  & 14.8                         & 14.9                         & 13.4                         & 18.5                         & 15.8          \\
male only           & 14.1          & \cellcolor[HTML]{EEEEEE}17.7 & \cellcolor[HTML]{EEEEEE}14.7 & 17.1                                  & 14.5                         & \textbf{13.4}                & 13.0                         & 19.1                         & 15.6          \\
mixed genders       & 14.0          & \cellcolor[HTML]{EEEEEE}17.0 & \cellcolor[HTML]{EEEEEE}14.6 & 16.4                                  & 14.4                         & 13.5                         & 13.0                         & 18.6                         & 15.4          \\ \hline
youngs only         & \textbf{13.7} & 16.3                         & 14.4                         & \cellcolor[HTML]{EEEEEE}\textbf{16.0} & \cellcolor[HTML]{EEEEEE}13.8 & \cellcolor[HTML]{EEEEEE}14.2 & 12.7                         & 17.5                         & 15.0          \\
middle-aged only    & 13.9          & 17.3                         & 14.7                         & \cellcolor[HTML]{EEEEEE}16.9          & \cellcolor[HTML]{EEEEEE}14.5 & \cellcolor[HTML]{EEEEEE}13.5 & 12.9                         & 19.0                         & 15.5          \\
seniors only        & 14.6          & 18.5                         & 15.4                         & \cellcolor[HTML]{EEEEEE}17.8          & \cellcolor[HTML]{EEEEEE}14.8 & \cellcolor[HTML]{EEEEEE}14.5 & 13.8                         & 19.2                         & 16.3          \\
mixed ages          & 13.9          & 17.2                         & 14.8                         & \cellcolor[HTML]{EEEEEE}16.6          & \cellcolor[HTML]{EEEEEE}14.3 & \cellcolor[HTML]{EEEEEE}14.4 & 13.2                         & 18.5                         & 15.6          \\ \hline
native only         & 14.3          & 17.4                         & 15.1                         & 17.5                                  & 13.9                         & 14.2                         & \cellcolor[HTML]{EEEEEE}12.8 & \cellcolor[HTML]{EEEEEE}19.5 & 15.8          \\
accented only       & 14.9          & 17.8                         & 15.2                         & 17.1                                  & 14.5                         & 15.0                         & \cellcolor[HTML]{EEEEEE}14.0 & \cellcolor[HTML]{EEEEEE}17.6 & 15.9          \\
mixed accents       & 14.2          & 17.3                         & 14.9                         & 16.5                                  & 14.5                         & 14.5                         & \cellcolor[HTML]{EEEEEE}13.4 & \cellcolor[HTML]{EEEEEE}18.0 & 15.6          \\ \hline
high diversity set  & 13.8          & \textbf{16.1}                & \textbf{14.3}                & \textbf{16.0}                         & \textbf{13.6}                & 13.9                         & 12.8                         & \textbf{16.9}                & \textbf{14.8}
\end{tabular}
\end{table*}

%% file: tables/best_results.tex
\begin{table}[th]
  \caption{Results of models trained on a larger number of speakers. We report the mean WER across demographic attributes in the column \emph{mean}.}
  \label{tab:best_results}
  \centering
\begin{tabular}{l|c|c|c}
\textbf{Trainset}   & \textbf{Speakers} & \textbf{test} & \textbf{mean} \\ \hline
Reference           & 5,324             & 13.7          & 14.8                              \\
All spks from train & 6,756             & 13.4          & 14.4                              \\
train + dev         & 9,500             & \textbf{13.3} & \textbf{14.3}                     %\\ \hline
%Whole train / 50k   & 6,756             & 11.4          & 12.4                              \\
%Whole train / 100k  & 6,756             & \textbf{10.9} & \textbf{12.0}                    
\end{tabular}
  
\end{table}

%% file: sections/5.conclusion.tex
\section{Conclusion}

There is still a long way to go to obtain more fair speech recognition models. Initiatives like the Common Voice corpus are a first step toward the collection of big and diverse speech datasets.
Even though Common Voice has important shortcomings, it is still useful to build efficient ASR models. We demonstrate how speaker diversity can be more important than demographic diversity, and therefore encourage researchers and the general public to give their voices in a collective effort to further expand the dataset. In future work, we plan to expand our analysis using biased pre-trained models and explore the influence of prosodic attributes (e.g. speech rate) on speech recognition. We also project to dive deeper into the differences between male and female speech and test for instance the impact of the pitch.